\documentstyle[twocolumn,aps,epsf]{revtex}

\begin{document}
\draft
\title { Anti-flow of K$^0_s$ Mesons in
                        6~AGeV Au~+~Au Collisions  }
\author{P.~Chung$^{(1)}$, N.~N.~Ajitanand$^{(1)}$,
J.~M.~Alexander$^{(1)}$,
M.~Anderson$^{(5)}$, D.~Best$^{(3)}$,F.P.~Brady$^{(5)}$,
T.~Case$^{(3)}$,
W.~Caskey$^{(5)}$, D.~Cebra$^{(5)}$,J.L.~Chance$^{(5)}$,
B.~Cole$^{(10)}$,
K.~Crowe$^{(3)}$, A.~Das$^{(2)}$, J.E.~Draper$^{(5)}$,
M.L.~Gilkes$^{(1)}$,
S.~Gushue$^{(1,8)}$, M.~Heffner$^{(5)}$,A.S.~Hirsch$^{(6)}$,
E.L.~Hjort$^{(6)}$, L.~Huo$^{(12)}$, M.~Justice$^{(4)}$,
M.~Kaplan$^{(7)}$, D.~Keane$^{(4)}$, J.C.~Kintner$^{(11)}$,
J.~Klay$^{(5)}$,D.~Krofcheck$^{(9)}$,R.~A.~Lacey$^{(1)}$,
J.~Lauret$^{(1)}$, M.A.~Lisa$^{(2)}$,H.~Liu$^{(4)}$, Y.M.~Liu$^{(12)}$,
R.~McGrath$^{(1)}$, Z.~Milosevich$^{(7)}$,G.~Odyniec$^{(3)}$,
D.L.~Olson$^{(3)}$,S.~Y.~Panitkin$^{(4)}$, C.~Pinkenburg$^{(1)}$,
N.T.~Porile$^{(6)}$,G.~Rai$^{(3)}$, H.G.~Ritter$^{(3)}$,
J.L.~Romero$^{(5)}$,
R.~Scharenberg$^{(6)}$, L.~Schroeder$^{(3)}$,B.~Srivastava$^{(6)}$,
N.T.B~Stone$^{(3)}$, T.J.M.~Symons$^{(3)}$, T.~Wienold$^{(3)}$,
R.~Witt$^{(4)}$ J.~Whitfield$^{(7)}$,
L.~Wood$^{(5)}$, and W.N.~Zhang$^{(12)}$
                       \\  (E895 Collaboration) }
        
\address{$^{(1)}$Depts. of Chemistry and Physics, SUNY
          Stony Brook, New York 11794-3400\\
$^{(2)}$Ohio State University, Columbus, Ohio 43210\\
$^{(3)}$Lawrence Berkeley National Laboratory,Berkeley, California, 94720\\
$^{(4)}$Kent State University, Kent, Ohio 44242\\
$^{(5)}$University of California, Davis, California, 95616\\
$^{(6)}$Purdue University, West Lafayette, Indiana, 47907-1396\\
$^{(7)}$Carnegie Mellon University, Pittsburgh, Pennsylvania 15213\\
$^{(8)}$Brookhaven National Laboratory, Upton, New York 11973\\
$^{(9)}$University of Auckland, Auckland, New Zealand\\
$^{(10)}$Columbia University, New York, New York 10027\\
$^{(11)}$St. Mary's College, Moraga, California  94575\\
$^{(12)}$Harbin Institute of Technology, Harbin, 150001 P.~R. China\\
}
%
%
\maketitle
\newpage
\begin{abstract}
%
We have measured the sideward flow of neutral
strange ($K^0_s$) mesons in 6~AGeV Au~+~Au collisions.
A prominent anti-flow signal is observed for an impact parameter
range (b~$\lesssim 7$~fm) which spans central and mid-central events.
Since the $K^0_s$ scattering cross section is relatively small in
nuclear matter, this observation suggests that the in-medium kaon vector
potential plays an important role in high density nuclear matter.
\end{abstract}

\narrowtext


	The production and properties of strange particles can provide
an important probe of hot and dense nuclear matter\cite{Senger98,Bass98}.
They can give insight into the fundamental aspects of chiral symmetry
restoration at high baryon density and/or temperature as well as information
relevant to neutron stars\cite{GBrown94,GQli97,Thorsson94,JWharris96}.
Recent theories of kaon propagation in nuclei all predict an important
role for the in-medium kaon-nucleon
potential\cite{Kaplan86,Brown91,Wass96,Schaffner96,Lutz98}.
This potential is comprised of two parts: (1) an attractive s-wave scalar
interaction, which can be linked to chiral symmetry breaking via the
mass of the strange quark, and (2) a vector potential which is thought
to be repulsive for kaons but attractive for anti-kaons.

	The repulsive kaon-nucleon interaction can lead to a net repulsion
of kaons away from nucleons resulting in the anti-flow of the
kaons\cite{gqli95}. In fact, it has been argued that
the sideward and elliptic flow pattern of kaons in relativistic
heavy ion reactions, can provide a good probe for the influence
of both the vector and scalar components of the kaon
potential\cite{gqli95,Wang98}.
A recent measurement has indicated essentially zero sideward flow
for the $K^+$s and $K^0_s$'s produced in 1.93 AGeV $^{58}$Ni~+~$^{58}$Ni
collisions. This result has been attributed to the repulsive
nature of the kaon-nucleon potential\cite{JRitman95}.
However, an alternative interpretation involving rescattering
effects has been shown to account for the data\cite{CDavid98}.
The observation of an out-of-plane or negative elliptic flow of $K^+$
mesons in 1~AGeV Au~+~Au collisions has been associated with the
repulsive part of the $K^+N$ potential\cite{YShin98}. Nonetheless,
a general consensus on the nature of the kaon-nucleon potential has
not been reached.

	Flow measurements for $K^0_s$ mesons can serve
as a unique probe of the kaon-nucleon potential in that they avoid
possible complications which could result from the Coulomb interaction
between a charged kaon and the associated emitting system.
Such measurements have been sparse because the relatively short
life-time $\tau$, of the $K^0_s$ (c$\tau$=2.68~cm) imposes the requirement
of a large acceptance device with high detection efficiency
and very good momentum resolution. In this letter we report on
the use of such a
device -- the E895 Time Projection Chamber~(TPC)\cite{GRai90} -- to make
detailed  measurements of the flow of neutral strange $K^0_s$
mesons in 6~AGeV Au~+~Au collisions. The data from these measurements
show prominent signatures for the anti-flow of $K^0_s$ for both
central and mid-central collisions, suggesting significant
in-medium effects for these particles.
%
The kaon potential, rather than that of the antikaon, is expected
to dominate in the present study,
since Relativistic Quantum Molecular Dynamics
(RQMD)\cite{Sorge95} calculations suggest an antikaon $\bar{K^0}$
contribution of $\lesssim 10$~\% in our data sample.
Preliminary results from this work have been reported\cite{PChung98}.

        The measurements have been performed with the E895 detector
system at the Alternating Gradient Synchrotron at the
Brookhaven National Laboratory. This system includes a TPC\cite{GRai90} and a
multi-sampling ionization chamber~(MUSIC)\cite{GBauer97}.
Details on the detector system have been reported
earlier\cite{CPinkenburg98,DBest97}.
The data presented here reflect the excellent coverage,
continuous 3D-tracking, and particle identification capabilities of the TPC.
All are important for the efficient detection and reconstruction of the
neutral strange $K^0_s$ mesons.

        The $K^0_s$'s have been reconstructed
from their pion decay-daughters
$K^0_s \longrightarrow \pi^+ + \pi^-$(branching
ratio $\sim$ 69\%) using the following procedure. First, all
TPC tracks in an event were reconstructed along with the
calculation of an overall event vertex. Second, each $\pi^+\pi^-$
pair was considered and their point of closest approach calculated.
$\pi^+\pi^-$ pairs whose trajectories intersected at a point other than
the main vertex were then evaluated to yield invariant mass $m_{inv}$
and momentum. All of these hypothesized $K^0_s$'s with
$0.4 < m_{inv} <0.6$ GeV/c$^2$ were then passed through a fully
connected feedforward multilayered neural network\cite{mjustice97}
trained to separate "true" $K^0_s$'s from the combinatoric background. The
network was trained from a set consisting of "true" $K^0_s$'s and
a set consisting of combinatoric background respectively.
"True" $K^0_s$'s were generated by
tagging and embedding simulated $K^0_s$'s in raw data events in
a detailed GEANT simulation of the TPC. The background or "fake"
$K^0_s$'s were generated via mixed events in which the
candidate daughter particles of the $K^0_s$ ($\pi^+ \pi^-$) were
chosen from different data events.

\begin{figure}
\centerline{\epsfysize=4.0in \epsffile{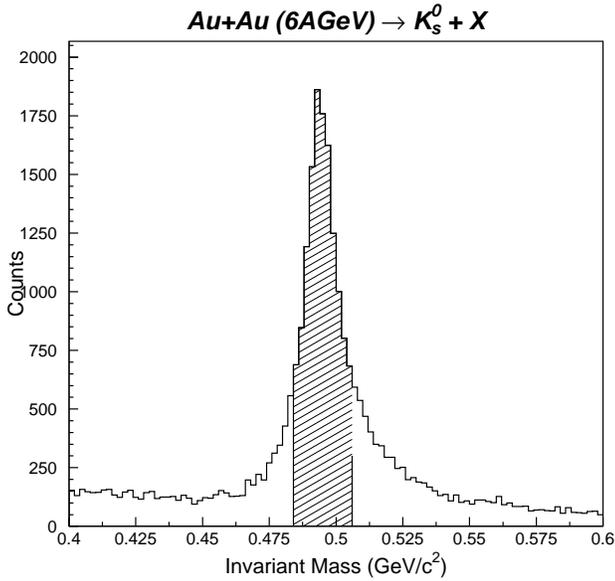}}
\vspace*{.6in}
\caption{   (a) Invariant mass distribution for $K^0_s$.
The hatched area indicates accepted $K^0_s$ particles.
}

\label{fig1}
\end{figure}

\begin{figure}
\centerline{\epsfysize=3.8in \epsffile{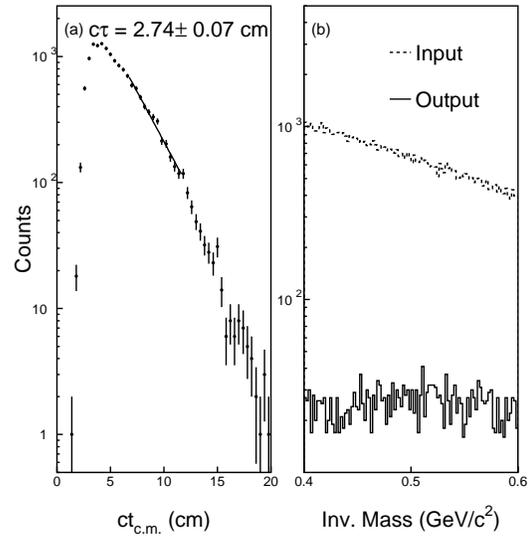}}
\vspace*{.4in}\caption{   (a) Decay-length distribution in the
	rest frame of the $K^0_s$. The solid curve is an exponential
	fit to the data (see text). (b) Invariant mass distributions
	for combinatoric events processed by the neural
	network (see text). The input
	and output distributions are represented by the dotted and
	solid curve respectively.
}

\label{fig2a}
\end{figure}

        The resulting experimental invariant mass distribution
for $K^0_s$'s is shown in Fig.~\ref{fig1}.
The distribution has been obtained for central and mid-central
events (charged particle multiplicity $\ge 100$) in which one or more
$K^0_s$'s have been detected. This multiplicity distribution
is estimated to correspond to an impact-parameter
range, b~$\lesssim 7$~fm. The distribution
shown in Fig.~\ref{fig1} is clearly peaked at the invariant mass
expected for the $K^0_s$ ($\sim .5$~GeV), and the excellent peak
to background ratio clearly demonstrates the reliability
with which the neural network is able to separate real $K^0_s$'s from
the combinatoric background. Fig.~\ref{fig2a}a,
shows the uncorrected (for detection efficiency) decay-length
distribution in the rest frame of the
$K^0_s$'s, i.e., $d_i/\gamma_i\beta_i$, where $d_i$ is the laboratory
distance from the main vertex to the decay vertex of each $K^0_s$, i. The
distribution is shown for the same $K^0_s$'s used to construct
the invariant mass distribution shown in Fig.~\ref{fig1}.
The distribution is characterized by a prominent
exponential tail. The apparent deficit below ct~$\sim$~4 cm
reflects the difficulty of $K^0_s$ reconstruction in the region of
high track density near to the main event vertex.
An exponential fit to the data over a region for which the
detection efficiency was determined to be constant (7 - 11 cm),
yields a c$\tau$ value of $2.74\pm 0.07$~cm.
This value is close to the expected
value of 2.68~cm and therefore serves as further confirmation
that the neural network does indeed separate true $K^0_s$'s
from the combinatoric background.
The reconstruction procedure produces a raw yield of
$\sim .03$ $K^0_s$'s per event.

	It is important to establish that the procedure used to
train the neural network does not lead to the spurious "creation"
of $K^0_s$'s. To do this, we have processed ``fake" $K^0_s$'s
through the neural network. Fig.~\ref{fig2a}b shows input
(dashed histogram) and output (solid histogram) invariant
mass distributions for ``fake" $K^0_s$'s generated via the mixed
event procedure discussed above. The absence of a peak (at
the $K^0_s$ mass) in the output distribution demonstrates that
the neural network is properly trained and does not "create"
spurious $K^0_s$'s.

        Our flow analysis, which follows the standard
transverse momentum analysis technique of Danielewicz and
Odyniec\cite{daniel85}, employs a gate, $0.485 \le m_{inv} \le 0.505$ GeV,
centered on the invariant mass peak to ensure a relatively
pure sample ($\sim$ 90~-~92\%)  of $K^0_s$'s ($\sim$ 12060 $K^0_s$'s above
background). The gate is represented by the hatched
area shown in Fig.~\ref{fig1}.
The orientation of the reaction plane vector ${\bf Q}$ was determined
for each event i, by summing over protons and light nuclei (j),
with charge $Z \le 2 $ in events associated with a selected
impact-parameter range\cite{pion_note};
%
$
{\bf Q}_i = \sum_{j}^n{w(y_j) {\bf p}^t_j/| {\bf p}^t_j|}.
$
%
Here, ${\bf p}^t_j$ and $y_j$ represent the transverse momentum vector and
rapidity, respectively for baryon~j . The weight $w(y_j)$ is assigned
the value $<p^x>/<p^t>$ where $<p^x>$ is the transverse momentum
in the reaction plane for baryons. $<p^x>$ is obtained from the
first pass of an iterative procedure. The orientation of the impact
parameter vector is random. Therefore, the distribution of the
determined reaction plane should be uniform (flat). We have established that
deviations from this uniformity are the direct result of deficiencies
in the acceptance of the TPC and have applied rapidity and
multiplicity dependent corrections following Ref.~\cite{CPinkenburg98}.
Such corrections ensure the absence of spurious flow signals which
could result from distortions in the reaction plane distribution.
The dispersion of the reaction plane, $<\phi_{12}>/2$
is estimated via the the sub-event method\cite{daniel85} to be
$\sim~37^o$ and $\sim~33^o$ for central and mid-central events
respectively.

\begin{figure}
\centerline{\epsfysize=3.8in \epsffile{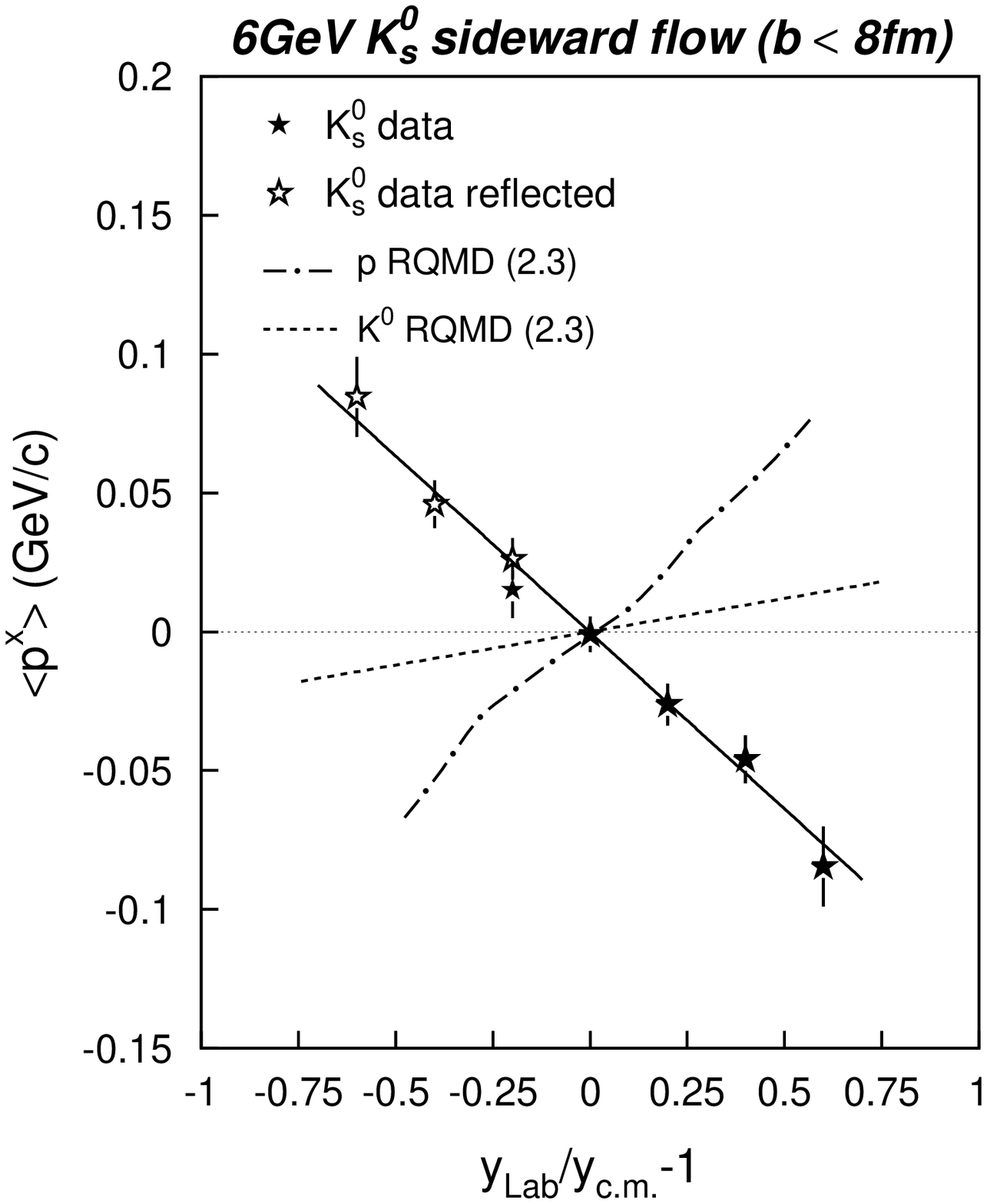}}
\vspace*{.4in}\caption{   Experimental $<p^x>$ vs $y_0$  for $K^0_s$ mesons
   (stars).  $<p^x>$ values have been corrected for reaction plane dispersion.
   The solid curve represents a linear fit to the data. Error
   bars are statistical. The dashed-dot and dotted curves represent
   results obtained for protons and $K^0_s$'s
   from RQMD~v~(2.3) (with mean-field). The RQMD results have been obtained
   for the same impact parameter and $p^t$ selection as that for the data.
}

\label{fig3}
\end{figure}

 The mean transverse momenta $<p^x>$, of
$K^0_s$'s in the reaction plane are shown as a function of
the normalized c.m rapidity, $y_0$ in Fig.~\ref{fig3}. Here
$y_0 = y_{Lab}/y_{cm} -1$; $y_{Lab}$ is the rapidity of the
emitted particle in the Lab, and $y_{cm}$ is the rapidity of the
c.m. The $<p^x>$ value shown for each rapidity bin (filled stars)
in Fig.~\ref{fig3} includes a small correction for, (a) the effect of
the $\sim$~8-10\% combinatoric background and (b) inefficiencies
associated with the detection of the $K^0_s$.
The correction for the combinatoric background has been made by evaluating
the $<p^x>$ [for each rapidity bin] for the experimental combinatoric
background followed by a weighted subtraction of these values from
the $<p^x>$ values obtained for the invariant mass selection
indicated in Fig.~\ref{fig1}. It is important to note here that the
magnitude of the flow for the combinatoric background is on average
$\sim$~4-5 times smaller than that for the $K^0_s$'s.
Thus, the net effect of the background would be an apparent
reduction in the flow.  The procedure
employed to evaluate the flow is as follows.
First, for each rapidity selection, we evaluate the $<p^x>$
as a function of $p^t$ over the range~0~-~0.7 GeV/c. For this
$p^t$ range the $<p^x>$ shows a linear dependence [on $p^t$] with a negative
slope. Following this evaluation,  we determine and correct the
$K^0_s$ $p^t$ distributions for the same rapidity selection.
Estimates for these efficiency corrections have been obtained by
tagging and embedding simulated $K^0_s$'s [with a flat input $p^t$
distribution for each y]
in raw data events in a detailed GEANT simulation of the TPC.
Subsequently, a weighted average (obtained by folding the
corrected $p^t$ distribution with the $p^t$-dependence of $<p^x>$ for
that y bin) was performed to obtain the $<p^x>$ as a function of
rapidity selection. It is noteworthy that this procedure takes account
of an $\sim$ 10\% correction to the flow resulting from the detection
efficiency for $K^0_s$'s.

The representative $<p^x>$ values shown in Fig.~\ref{fig3} are
for $p^t \le 700$ MeV/c, and b~$\lesssim 7$~fm. The limited acceptance
for $K^0_s$ detection at negative rapidities results in an apparent
cut-off for $y_0 \le -0.30$. The $<p^x>$ values clearly follow
an anti-flow pattern. Similarly prominent antiflow patterns have been
obtained for more central and less central events.
A correction factor of 1.44, has been applied to the data
shown in Fig.~\ref{fig3} to account for the reaction plane
dispersion\cite{dan87,oll97,Postkanzer98}. A linear fit to
these data yields a slope of $-127 \pm 20$ MeV/c.
The dotted and dot-dashed curves shown in Fig.~\ref{fig3}
represent flow results obtained from RQMD\cite{Sorge95} calculations
for $K^0_s$'s and protons respectively. The calculations, which
have been performed for the same impact parameter and $p_t$ range as that
for the data, include the effects of a mean-field as well as
rescattering. However, they do not include the kaon-nucleon
potential. Fig.~\ref{fig3} shows calculated trends for protons
which are in qualitative agreement with the data\cite{hliu97}.
By contrast, the experimental flow pattern observed for $K^0_s$'s
is clearly at odds with the results obtained from the calculations.
This difference is particularly striking for the comparison of both
the magnitude and sign of the flow. Here, it is important to stress
that unlike pions, $K^0_s$'s have a long mean free path in nuclear matter
due to their rather small scattering cross section
($\sigma_{K^0+p} \sim 10$~mb and $\sigma_{\pi +p} \sim 100$~mb ).
This being the case, one cannot account for the antiflow of $K^0$'s
via the reabsorption mechanism commonly exploited to explain pion
anti-flow\cite{Oeschler}. Thus, we attribute the disagreement between data
and theory to the absence of an appropriate kaon-nucleon potential
in RQMD, and conclude that the experimentally observed flow pattern
[for $K^0_s$'s] is more consistent with predictions for a
strong influence of the repulsive kaon vector
potential in the nuclear medium\cite{gqli95,Wang98}.


In sum, we have measured the transverse flow of neutral
strange $K^0_s$ mesons in central and mid-central 6AGeV
Au+Au collisions. The data show a clear anti-flow signal which
is in stark contrast to that observed for protons at the same beam energy.
This contrast is more pronounced than for lower energies and
a smaller system size\cite{JRitman95},
possibly because the effects of the repulsive kaon mean field
become more significant with the higher baryon densities
expected at 6~AGeV. Our sidewards flow data, while apparently
different from those for the $^{58}$Ni~+~$^{58}$Ni
system (1.93 AGeV)\cite{JRitman95}, are not incompatible with the
conclusions drawn about the role of the vector potential.


        This work was supported in part by the U.S. Department
of Energy under grants DE-FG02-87ER40331, DE-FG02-89ER40531,
DE-FG02-88ER40408, DE-FG02-87ER40324, and contract
DE-AC03-76SF00098; by the US  National
Science Foundation under Grants  PHY-98-04672, PHY-9722653,
PHY-96-05207,PHY-9601271, and INT-9225096; by the University of
Auckland Research Committee, NZ/USA Cooperative Science Programme
CSP 95/33; and by the National Natural Science Foundation of P.R. China
under grant 19875012.



%



\begin{references}
\bibitem{Senger98} C. M. Ko and G. Q. Li, J. Phys. G22, 1673 (1996)
\bibitem{Bass98} S.~A.~Bass, et al., J.Phys. G25, R1, (1999)
\bibitem{GBrown94} G. E. Brown et al. Nucl. Phys. A567, 937 (1994)
\bibitem{GQli97} G. Q. Li et al., Nucl. Phys. A625, 372 (1997)
\bibitem{Thorsson94} V. Thorsson et al., Nucl. Phys. A572, 693 (1994)
\bibitem{JWharris96} J. W. Harris et al., Annu. Rev. Part. Sci. 46, 71 (1996)
\bibitem{Kaplan86} D. B. Kaplan et al., Phys. Lett. B175, 57 (1986)
\bibitem{Brown91} G. E. Brown et al., Phys. Rev. C43, 1881, (1991)
\bibitem{Wass96} T. Waas et al., Phys. Lett. B379, 34 (1996)
\bibitem{Schaffner96} J. Schaffner et al., Phys. Rev. C53, 1416 (1996)
\bibitem{Lutz98} M. Lutz, Phys. Lett. B 426, 12 (1998)
\bibitem{gqli95} G.Q. Li, et al.,  Phys. Rev. Lett. 74, 235 (1995),
                 Bao-An Li et al., Phys. Rev. {\bf C}60, 3492, (1999)
\bibitem{Wang98} Z. S. Wang, et al., nucl-th/9809043, (1998)
\bibitem{JRitman95} J.~Ritman et al., Z Phys. A352, 355, (1995)
\bibitem{CDavid98} C.~David et al., Nuc. Phys. A nucl-th/9805017
\bibitem{YShin98} Y. Shin et al., Phys. Rev. Lett. 81, 1576 (1998)
\bibitem{GRai90} G. Rai et al., IEEE Trans. Nucl. Sci. 37, 56 (1990)
\bibitem{Sorge95}H.\ Sorge Phys. Rev. {\bf C}52, 3291, (1995)
%
\bibitem{PChung98} P.~Chung et al., (E895 Collaboration)
		Journal of Phys. G., 25, 255 (1999)
\bibitem{GBauer97} G. Bauer et al., NIM A{\bf  386},  249 (1997)

\bibitem{CPinkenburg98} C.~Pinkenburg et al. (E895 Collaboration),
  Phys. Rev. Lett. 83, 1295 (1999)

\bibitem{DBest97} D.~Best for the E895 Collaboration,

                 J. Phys. G:~Nucl. Part. Phys. 23, 1873, (1997)

\bibitem{mjustice97} M. Justice, Nucl. Instrum. Methods
        Phys. Res. A 400, 463 (1997)


\bibitem{daniel85} P. Danielewicz and G. Odyniec, Phys. Lett. B157,146
(1985)

\bibitem{pion_note} Unique separation of $\pi^+$'s and protons was not
achieved for all rigidities. Consequently, a small fraction of $\pi^+$'s
were included in the sample used to determine the reaction plane.
The small influence of this pion contamination on the extracted flow
value is accounted for via the dispersion correction for the reaction plane.


\bibitem{dan87}

P.\ Danielewicz {\em et al.}, Phys.\ Rev.\ C {\bf 38}, 120,(1988).

\bibitem{oll97} J.-Y.\ Ollitrault, nucl-ex/9711003 v2.
%

\bibitem{Postkanzer98} A. M. Poskanzer and S. A. Voloshin,
  Phys. Rev {\bf C}58, 1671, (1998)

\bibitem{hliu97} H. Liu et al. (E895 Collaboration)

{\it Quark Matter '97, Proc.\ 13th Int.\
Conf.\ on Ultra--Relativistic Nucleus--Nucleus Collisions,
Tsukuba, Japan, 1997,} ed.\ P.\ T.~Hatsuda {\em et al.},
Nucl. Phys. {\bf A638}, 451c (1998), H. Liu et al. (E895 Collaboration)
Phys. Rev Let. (submitted).

\bibitem{Oeschler} H. Oeschler, IKDA 98/20 Report, (1998).

\end{references}
\end{document}